# Quantum resolution sizes and atomic bonding states of two-dimensional SnO

Yu Wang, Yunhu Zhu, Yixin Li, Maolin Bo*

Key Laboratory of Extraordinary Bond Engineering and Advanced Materials Technology (EBEAM) of Chongqing, Yangtze Normal University, Chongqing 408100, China

* Corresponding author Email: bmlwd@yznu.edu.cn

**Abstract**

Understanding the interatomic bonding and electronic properties of two-dimensional (2D) materials is essential for developing high-performance 2D semiconductor materials. In this study, we calculated the band structure, electronic properties, and bonding characteristics of SnO in 2D materials using density functional theory alongside bond energy and bond charge models. Our analysis of atomic bonding provided a detailed examination of the interatomic bonding and charge transfer within the layered SnO structure. The findings significantly advanced the understanding of local bonding states on the surfaces of 2D structural materials. Furthermore, a renormalization method was employed to manipulate energy levels and determine wave function at varying quantum resolutions. This approach is particularly valuable for describing the size-dependent properties and phase transitions in nanomaterials.

## 1. Introduction

Two-dimensional (2D) materials, owing to their unique physical and chemical properties, have demonstrated significant potential in various fields, including electronics, optics, and energy applications [1-3]. In recent years, 2D SnO has garnered significant attention. Unlike traditional van der Waals materials, the interlayer bonding in SnO involves metal bonds or electrostatic interactions, resulting in strong electronic coupling [3, 4]. The band gap of SnO can be adjusted over a broad range, ranging from 0.60 eV in the infrared spectrum for bulk samples to 3.65 eV in the ultraviolet spectrum for monolayer samples [5].

SnO exhibits significant advantages as an electrode material, with its high electron mobility making it a promising candidate for efficient electronic devices [6, 7]. Furthermore, its semi-metallic properties are critical in spintronics, where complete spin-polarized semimetallicity can be achieved through hole doping [8, 9]. This enables the design and production of efficient spin filters and other devices. Additionally, the tunable band gap of SnO allows for the optimization of its semi-metallic properties to meet different needs. Beyond applications in electrode materials and spintronics, SnO holds promise in other fields such as photocatalysis, sensing, quantum computing, and biomedicine [10-15]. These diverse applications make SnO a versatile material for in-depth exploration and research.

This study utilized density functional theory (DFT) calculations to conduct in-depth research of the band structure, density of states, and energy renormalization properties of 2D SnO materials. The electronic properties and bonding states of 2D SnO were calculated, revealing the bonding characteristics at both the electronic and atomic levels. This research provides a theoretical foundation for understanding the intrinsic properties of 2D SnO, while also offering insights into the design and optimization of related 2D materials. Furthermore, exploring these characteristics supports the advancement of 2D material development and lays a strong groundwork for the development of new materials and technologies.

## 2. Methods

### 2.1 DFT Calculation

The atomic bonding, electronic properties, and structural relaxation of bilayer SnO materials were investigated using the Cambridge sequential total energy package, which employed total energy calculations with a plane-wave pseudopotential approach. This analysis focused on the bonding characteristics, structural stability, energetics, and electronic properties of bilayer SnO materials. We utilized the hybrid density functional HSE06 to describe electron exchange and correlation potentials [16]. Additionally, the Tkatchenko-Scheffler scheme was employed for DFT dispersion corrections to account for long-range van der Waals interactions. All the structures were fully optimized without symmetry constraints until the forces were reduced to less than 0.01 eV/Å and the energy tolerances were lower than $5.0 \times 10^{-6}$ eV per atom. A convergence threshold of $1.0 \times 10^{-6}$ eV/atom was used for self-consistent field calculations. A vacuum space of at least 14 Å in the direction normal to the bilayer was used to prevent interactions between periodic images. The cutoff energies, bandgaps, and k-point grids are shown in **Table 1**.

**Table 1** Cut-off energies, band gaps, and *k*-points of SnO

| Element | Cut-off energy | *k*-point | Band gap (HSE06) |
|---|---|---|---|
| SnO(α) | 750 eV | 10×10×2 | 1.508 eV |
| SnO(β) | 750 eV | 10×10×2 | 1.090 eV |

**2.2 Renormalization group method**

Based on the Kardar-Parisi-Zhang equation [17], the functional integral, or path integral, is expressed as

$$Z(\Lambda) = \int_{\Lambda} \mathcal{D}h e^{-S(h)} \tag{1}$$

Where the action is

$$S(h) = \frac{1}{2}\int d^D x dt \left[ \frac{\partial h}{\partial t} - \nabla^2 h - \frac{g}{2}(\nabla h)^2 \right]^2 \tag{2}$$

**Eq. 2** defines a field theory. Similar to other field theories, cutoff parameter $\Lambda$ must be introduced. To achieve this, we integrated part of the field configurations $h(\boldsymbol{x}, t)$ separately. All Fourier components $\boldsymbol{k}$ and $\omega$ of these configurations satisfied being no greater than a certain value Λ. In principle, since this is a non-relativistic theory, $\boldsymbol{k}$ and another quantity b must each have a cutoff value. For simplicity, these cutoffs are collectively referred to as Λ.

From a physical perspective, since the random driving term, $\eta(x, t)$, represents white noise, it indicates that values at different positions $\boldsymbol{x}$ and $\boldsymbol{x'}$ (and similarly at different times) are uncorrelated. On a microscopic scale, this resulted in a highly uneven surface, as shown in **Fig. 1(a).** However, this study focused on physical phenomena at a larger scale, *L*. To observe these phenomena, we adopted a "blurred" perspective, as if viewing the surface through a less-detailed lens. At this scale, the surface appeared smoothed, as shown in **Fig. 1(b)**.

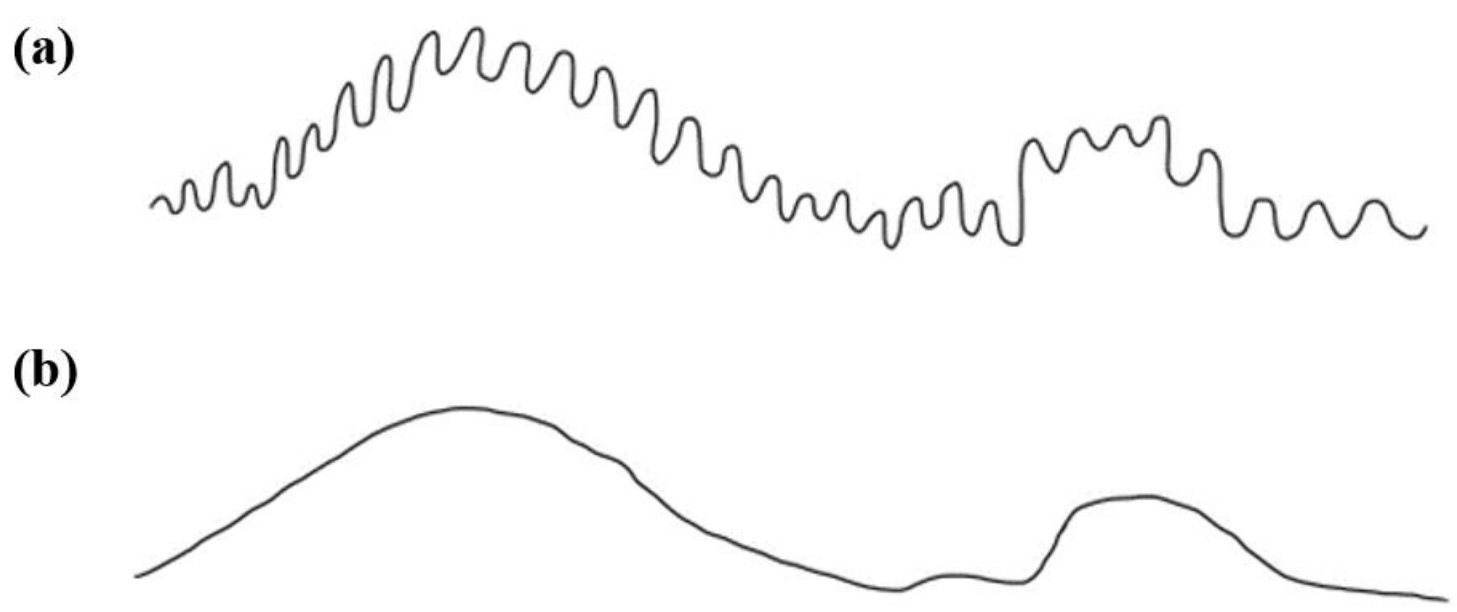

**Fig. 1** White noise across different scales (a) Small scale (b) Large scale

The renormalization group constitutes a formal framework that connects different scales, specifically different energy thresholds. To implement this concept physically, the length scale can be modified in the functional integral in **Eq. 1**. In Euclidean space $\lambda\varphi^4$ theory [18]:

$$Z(\Lambda) = \int_{\Lambda} \mathcal{D}\varphi e^{-\int d^d x \mathcal{L}(\varphi)} \tag{3}$$

Among them, symbol $\int_{\Lambda}$ indicates an integration that is limited to field configurations where *φ(x)* is defined by the following integral: $\varphi(x) = \int \left[ d^d k / (2\pi)^d \right] e^{ikx} \varphi(k)$.

Function $\varphi(k)$ is zero for any $k$ whose magnitude, given by $|k| \equiv \left(\sum_{i=1}^{d} k_i^2\right)^{\frac{1}{2}}$, exceeds cutoff $\Lambda$. The system's resolution is inversely related to $\Lambda$, with $L=1/\Lambda$.

Let $\Lambda \to \Lambda - \delta\Lambda$ (with $\delta\Lambda > 0$) introduce a degree of blurring. Field $\varphi$ is decomposed into $\varphi = \varphi_s + \varphi_w$, where $s$ represents the smooth component and $w$ represents the wiggly component. Fourier components $\varphi_s(k)$ and $\varphi_w(k)$ are non-zero only within specific ranges of $k$; $\varphi_s(k)$ is non-zero for $|k| \le q(\Lambda - \delta\Lambda)$, and $\varphi_w(k)$ is non-zero for $(\Lambda - \delta\Lambda) \le q|k| \le q\Lambda$. Substituting these values into **Eq. 3**, the following expression is obtained:

$$Z(\Lambda) = \int_{\Lambda-\delta\Lambda} \mathcal{D}\varphi_s e^{-\int d^d x \mathcal{L}(\varphi_s)} \int \mathcal{D}\varphi_w e^{-\int d^d x \mathcal{L}_1(\varphi_s,\varphi_w)}$$

(4)

In expression $\mathcal{L}_1(\varphi_s,\varphi_w)$, all terms depend on $\varphi_w$. Integrating over $\varphi_w$, the result is

$$e^{-\int d^d x \delta\mathcal{L}(\varphi_s)} \equiv \int \mathcal{D}\varphi_w e^{-\int d^d x \mathcal{L}_1(\varphi_s,\varphi_w)}$$

(5)

Thus, it follows that

$$Z(\Lambda) = \int_{\Lambda-\delta\Lambda} D\varphi_s e^{-\int d^d x \left[\mathcal{L}_1(\varphi_s) + \delta\mathcal{L}(\varphi_s)\right]}$$

(6)

At this point, formally speaking, the theory is now entirely described by the "smooth" field, $\varphi_s$.

Consider general expression $\mathcal{L} = \frac{1}{2}(\partial\varphi)^2 + \sum_n \lambda_n \varphi^n + \cdots$, where $\lambda_2$ corresponds to $\frac{1}{2}m^2$ and $\lambda_4$ is denoted as $\lambda$. The terms involving $\partial\varphi_s$ and $\partial\varphi_w$, among others, are integrated to zero, leading to the following:

$$\int d^d x \mathcal{L}_1(\varphi_s,\varphi_w) = \int d^d x \left(\frac{1}{2}(\partial\varphi_w)^2 + \frac{1}{2}m^2\varphi_w^2 + \cdots\right)$$

(7)

In the given context, $\varphi_s$ is included within $(\cdots)$. This theory describes the interaction of the $\varphi_w$ field with itself and the background field $\varphi_s(x)$.

From the perspective of symmetry, $\partial\mathcal{L}(\varphi_s)$ has the same form as $\mathcal{L}(\varphi_s)$, differing only in their coefficients. By adding $\partial\mathcal{L}(\varphi_s)$ to $\mathcal{L}(\varphi_s)$, coupling constants $\lambda_n$, and the coefficient of $\frac{1}{2}(\partial\varphi_s)^2$ are altered. These changes lead to a flow within the previously described space of the coupled functions.

Comparing **Eq. 6** with **Eq. 1**, replace $\int_{\Lambda-\delta\Lambda}$ in **Eq. 6** with $\int_{\Lambda}$, and introduce a real number $b<1$ such that $\Lambda-\delta\Lambda=b\Lambda$. In $\int_{\Lambda-\delta\Lambda}$, field integration is conducted for fields that satisfy $|k|\le b\Lambda$. Perform a simple variable substitution using $k=-bk'$, where $|k'|\le\Lambda$. We then used $x=x'/b$ to ensure $e^{ikx}=e^{ik'x'}$. By substituting these transformations, the following is obtained:

$$\int d^dx\mathcal{L}(\varphi_s)=\int d^dx'b^{-d}\left[\frac{1}{2}b^2(\partial'\varphi_s)^2+\sum_n\lambda_n\varphi_s^n+\cdots\right] \tag{8}$$

In the text, $\partial'=\partial/\partial x'=(1/b)\partial/\partial x$. Expression $\varphi'$ is defined as $b^{2-d}(\partial'\varphi_s)^2=(\partial'\varphi')^2$, which implies $\varphi'=b^{\frac{1}{2}(2-d)}\varphi_s$. Consequently, **Eq. 8** transforms into

$$\int d^dx\mathcal{L}(\varphi_s)=\int d^dx'\left[\frac{1}{2}(\partial'\varphi')^2+\sum_n\lambda_nb^{-d+(n/2)(d-2)}\varphi'^n+\cdots\right] \tag{9}$$

Therefore, the coefficient $\lambda'_n$ for $\varphi'^n$ can be written as

$$\lambda'_n=b^{(n/2)(d-2)-d}\lambda_n \tag{10}$$

where *n* indicates the number of fields interacting with each other, and *d* denotes the dimension.

### 2.3 Binding-energy and Bond-Charge(BBC) model

BBC models consist of three components: the binding energy model, bond charge model, and the Hamiltonian of the bond band model. According to the tight-binding approximation model of band theory [19, 20]:

$$E(\vec{k}) = \varepsilon_i - J_0 - \sum_{Rs} J(\vec{R}_s) e^{-i\vec{k}\cdot\vec{R}_s} \tag{11}$$

Among them, $J_0 = -\int \varphi *_i (\vec{r} - \vec{R}_m)(V(\vec{r}) - v_a(\vec{r} - \vec{R}_m))\varphi_i(\vec{r} - \vec{R}_m) d\vec{r}$ ,

$\varepsilon_i = \int \varphi *_i (\vec{r} - \vec{R}_m)(-\frac{\hbar^2}{2m}\nabla^2 + v_a(\vec{r} - \vec{R}_m))\varphi_i(\vec{r} - \vec{R}_m) d\vec{r}$ ,

$J(\vec{R}_n - \vec{R}_m) = -\int \varphi_i^* [\vec{\xi} - (\vec{R}_n - \vec{R}_m)][V(\vec{\xi}) - v(\vec{\xi})]\varphi_i(\vec{\xi}) d\vec{\xi}_m$ , $\vec{\xi} = \vec{r} - \vec{R}_m$ , $\vec{R}_s = \vec{R}_n - \vec{R}_m$ .

$J_0$ represents the energy level shift of $\varepsilon_i$ caused by the potentials of the (N-1) atoms other than the lattice point in the crystal lattice.The binding energy shifts of the bulk and surface atoms can be expressed as follows:

$$E_V(B) - E_V(0) = J_0 \tag{12}$$

Then, we have $\Delta E_v(i) = \gamma^m J_0$ . Furthermore, based on $\psi_{nk}(\vec{r}) = \frac{1}{\sqrt{N}} \sum_{Rm} e^{i\vec{k}\cdot\vec{R}_m} \varphi_n(\vec{r} - \vec{R}_m)$ , $V_{cry}(\vec{r}) = V(\vec{r}) - v_a(\vec{r} - \vec{R}_m)$ , and $\gamma V_{cry}(\vec{r}) = -\gamma \sum_{i,j,\vec{R}_j \neq 0} \frac{1}{4\pi\varepsilon_0} \frac{Z' e^2}{|\vec{r}_i - \vec{R}_j|}$ . For atomic bonding, **Eq. 12** can be written as:

$$\gamma^m = \frac{E_V(i) - E_V(0)}{E_V(B) - E_V(0)} = \left(\frac{E_V(x) - E_V(0)}{E_V(B) - E_V(0)}\right)^m \approx \left(\frac{Z_x d_b}{Z_b d_x}\right)^m = \left(\frac{Z_b - \mu_v}{Z_b}\right)^m \left(\frac{d_x}{d_b}\right)^{-m} = \left(\frac{d_x}{d_b}\right)^{-m'} \approx \left(\frac{d_x}{d_b}\right)^{-m}$$

In $m' = m\left(1 - \frac{\ln \frac{Z_b - \mu_v}{Z_b}}{\ln\left(\frac{d_x}{d_b}\right)}\right)$, $\mu_v$ is very small; therefore, $\frac{Z_b - \mu_v}{Z_b} \approx 1$ and $m \approx m'$ were used. For compounds, $m \neq 1$.

The Hamiltonian for the system in the BC model is expressed as follows **(see supplemental material)**:

$$H = \sum_{k\sigma} \frac{\hbar^2 k^2}{2m} a_{k\sigma}^{\dagger} a_{k\sigma} + \frac{e_1^2}{2V} \sum_{q} {}^{*} \sum_{\vec{k}\sigma} \sum_{\overline{K'}\lambda} \frac{4\pi}{q^2} a_{\vec{k}+\vec{q},\sigma}^{\dagger} a_{\overline{K'}-\vec{q},\lambda}^{\dagger} a_{\overline{K'}\lambda} a_{\vec{k}\sigma}$$

(13)

When delving into the properties of the electronic system, by considering the dynamic changes of the electron Hamiltonian before and after relaxation, we can accurately apply the basic principles of perturbation theory. From the perspective of the microscopic mechanism, the electron Hamiltonian after relaxation can essentially be interpreted as having a perturbing effect on the charge distribution within the system. This perturbation, in turn, gives rise to a series of changes in the electronic states and related physical properties.The electron-interaction terms for density fluctuations, which are primarily caused by electrostatic shielding through electron exchange, are given by:

$$\begin{aligned}
\Delta H &= H_{ee}^{'} - H_{ee} \\
&= \frac{e_1^2}{2V} \sum_{\vec{k}} \sum_{\overline{K'}} \sum_{\vec{q}} \sum_{\sigma\lambda} \frac{4\pi}{q'^2} a_{\vec{k}+\vec{q},\sigma}^{\dagger} a_{\overline{K'}-\vec{q},\lambda}^{\dagger} a_{\overline{K'}\lambda} a_{\vec{k}\sigma} - \frac{e_1^2}{2V} \sum_{q} {}^{*} \sum_{\vec{k}\sigma} \sum_{\overline{K'}\lambda} \frac{4\pi}{q^2} a_{\vec{k}+\vec{q},\sigma}^{\dagger} a_{\overline{K'}-\vec{q},\lambda}^{\dagger} a_{\overline{K'}\lambda} a_{\vec{k}\sigma} \\
&= \frac{e_1^2}{2V} \sum_{\vec{k}} \sum_{\overline{K'}} \sum_{\vec{q}} \sum_{\sigma\lambda} \frac{4\pi}{q^2 + \mu^2} a_{\vec{k}+\vec{q},\sigma}^{\dagger} a_{\overline{K'}-\vec{q},\lambda}^{\dagger} a_{\overline{K'}\lambda} a_{\vec{k}\sigma} - \frac{e_1^2}{2V} \sum_{q} {}^{*} \sum_{\vec{k}\sigma} \sum_{\overline{K'}\lambda} \frac{4\pi}{q^2} a_{\vec{k}+\vec{q},\sigma}^{\dagger} a_{\overline{K'}-\vec{q},\lambda}^{\dagger} a_{\overline{K'}\lambda} a_{\vec{k}\sigma} \\
&= \frac{1}{4\pi\varepsilon_0} \sum_{i=1}^{N} \sum_{\substack{j=1 \\ i \neq j}}^{N} \frac{e^2}{2\left|\vec{r}_i - \vec{r}_j\right|} e^{-\mu\left(\vec{r}_i - \vec{r}_j\right)} - \frac{1}{4\pi\varepsilon_0} \sum_{i=1}^{N} \sum_{\substack{j=1 \\ i \neq j}}^{N} \frac{e^2}{2\left|\vec{r}_i - \vec{r}_j\right|}
\end{aligned}$$

(14)

Deformation bond energy $\delta V_{bc}$ can be represented as follows:

$$
\begin{aligned}
\Delta V_{bc} &= V_{e1} - V_{e2} \\
&= \sum_{i,j} \frac{1}{4\pi\varepsilon_0} \frac{e^2}{2\left|\vec{r}-\vec{r'}\right|} \int d^3r \int d^3r' \rho(\vec{r}) \rho(\vec{r'}) e^{-\mu(\vec{r}-\vec{r}')} - \sum_{i,j} \frac{1}{4\pi\varepsilon_0} \frac{e^2}{2\left|\vec{r}-\vec{r'}\right|} \int d^3r \int d^3r' \rho(\vec{r}) \rho(\vec{r'}) \\
&= \sum_{i,j} \frac{1}{4\pi\varepsilon_0} \frac{e^2}{2\left|\vec{r}-\vec{r'}\right|} \int d^3r \int d^3r' \delta\rho(\vec{r}) \delta\rho(\vec{r'})
\end{aligned}
$$

$$
\delta V_{bc} = \frac{1}{4\pi\varepsilon_0} \frac{e^2}{2\left|\vec{r}-\vec{r'}\right|} \int d^3r \int d^3r' \delta\rho(\vec{r}) \delta\rho(\vec{r'}) \tag{15}
$$

Also, $\delta\rho$ satisfies the following relationship:

$$
(\delta\rho_{Hole\text{-}electron} \le \delta\rho_{Antibonding\text{-}electron} < \delta\rho_{No\ charge\ tranfer} = 0 < \delta\rho_{Nonbonding\text{-}electron} \le \delta\rho_{Bonding\text{-}electron}) \tag{16}
$$

For atomic (strong) bonding states,

$$
\delta\rho_{Hole\text{-}electron}(\vec{r})\, \delta\rho_{Bonding\text{-}electron}(\vec{r}') < 0 (Strong\ Bonding). \tag{17}
$$

For atomic nonbonding or weak-bonding states,

$$
\begin{cases}
\delta\rho_{Hole\text{-}electron}(\vec{r})\, \delta\rho_{Nonbonding\text{-}electron}(\vec{r}') < 0 (Nonbonding\ or\ Weak\ Bonding) \\
\delta\rho_{Antibonding\text{-}electron}(\vec{r})\, \delta\rho_{Bonding\text{-}electron}(\vec{r}') < 0 (Nonbonding\ or\ Weak\ Bonding) \\
\delta\rho_{Antibonding\text{-}electron}(\vec{r})\, \delta\rho_{Nonbonding\text{-}electron}(\vec{r}') < 0 (Nonbonding)
\end{cases} \tag{18}
$$

For atomic antibonding states,

$$
\begin{cases}
\delta\rho_{Nonbonding\text{-}electron}(\vec{r})\, \delta\rho_{Bonding\text{-}electron}(\vec{r}') > 0 (Antibonding) \\
\delta\rho_{Hole\text{-}electron}(\vec{r})\, \delta\rho_{Antibonding\text{-}electron}(\vec{r}') > 0 (Antibonding) \\
\delta\rho_{Hole\text{-}electron}(\vec{r})\, \delta\rho_{Hole\text{-}electron}(\vec{r}') > 0 (Antibonding) \\
\delta\rho_{Antibonding\text{-}electron}(\vec{r})\, \delta\rho_{Antibonding\text{-}electron}(\vec{r}') > 0 (Antibonding) \\
\delta\rho_{Nonbonding\text{-}electron}(\vec{r})\, \delta\rho_{Nonbonding\text{-}electron}(\vec{r}') > 0 (Antibonding) \\
\delta\rho_{Bonding\text{-}electron}(\vec{r})\, \delta\rho_{Bonding\text{-}electron}(\vec{r}') > 0 (Antibonding)
\end{cases} \tag{19}
$$

The formation of chemical bonds is related to fluctuations in electron density $\delta\rho$.

## 3. Results and discussion

### 3.1 Geometry structure

The geometric structures of SnO(α) and SnO(β) are shown in **Fig. 2**. The Sn and O atoms are arranged in an alternating $Sn_{1/2}$−O−$Sn_{1/2}$ layered sequence along the [001] crystallographic direction. We optimized the geometric structures of SnO(α) and SnO(β) using DFT calculations. The optimized lattice parameters, bond lengths, bond angles, and interlayer distances are listed in **Table 2.** In the two different crystal structures of SnO(α) and SnO(β), variations in the atomic radii and electronegativities of Sn atoms and O atoms result in slight differences in interlayer distances. Specifically, the measured interlayer distance for SnO(α) is 2.792 Å, compared to 2.744 Å for SnO(β). Similarly, the Sn-O bond length for SnO(α) is 2.260 Å, which is marginally longer than the 2.256 Å observed in SnO(β). The bond angles of the two crystal forms are 118.01° and 118.09°, respectively. Additionally, we conducted a molecular dynamics simulation to evaluate the stability of SnO(α) and SnO(β). The simulation results indicate that both SnO(α) and SnO(β) exhibit excellent structural stability.Molecular dynamics simulation shows that SnO(α) and SnO(β) structure are stable. The results are shown in the **supplemental material**.

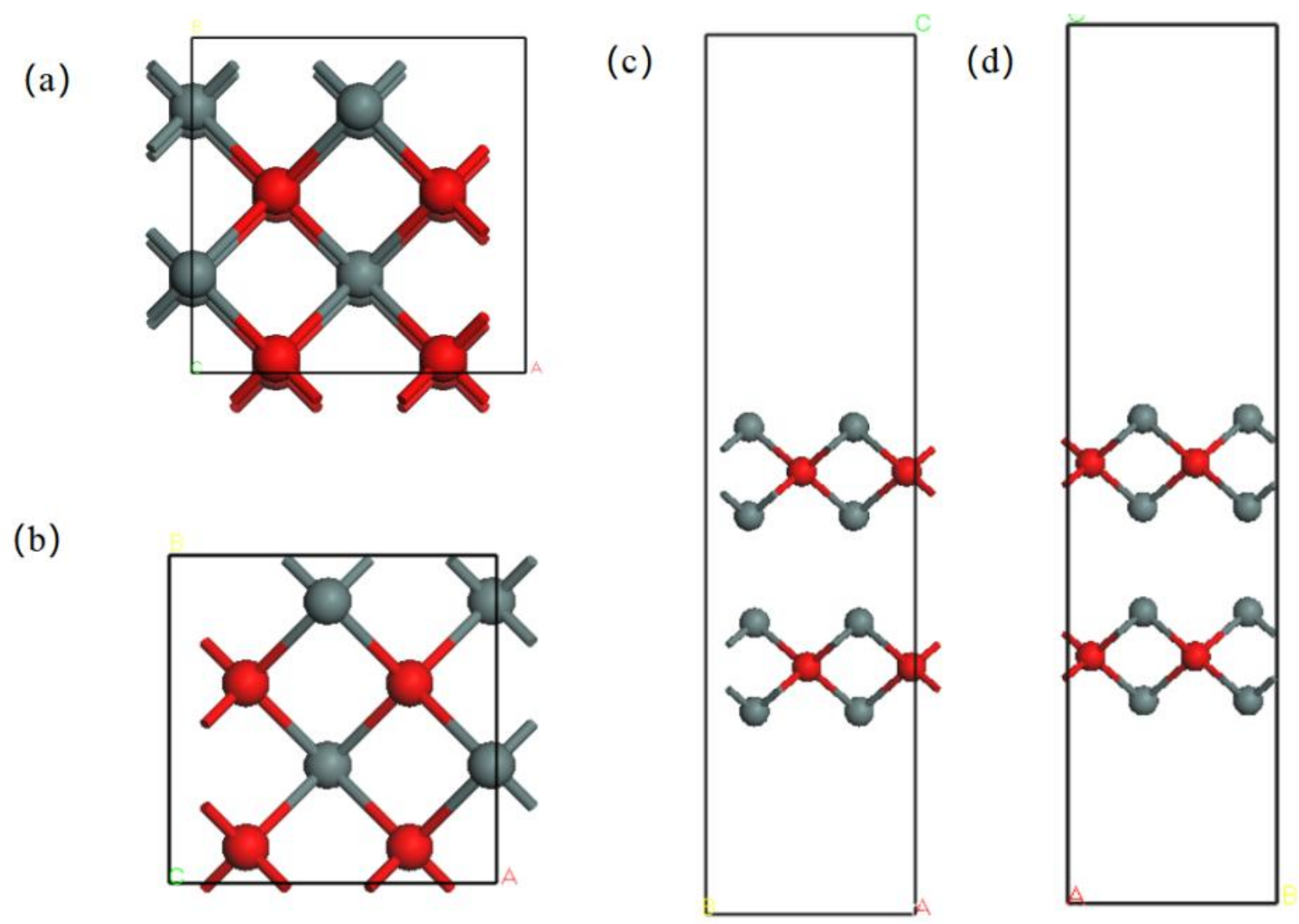

**Fig. 2** Crystal structures of SnO, with panels (a) and (c) showing SnO(α) and panels (b) and (d) displaying SnO(β).

**Table 2** Bond length, bond angle, **i**nterlayer distance and lattice parameters of 2D SnO(α) and SnO(β)

| | Interlayer distance | Lattice parameters（° / Å） | | | | | | Bond length | Bond angle |
|---|---|---|---|---|---|---|---|---|---|
| | | α | β | γ | a | b | c | | |

| | （Å） | | | | | | | （Å） | （°） |
|---|---|---|---|---|---|---|---|---|---|
| SnO(α) | 2.792 | 90 | 90 | 90 | 5.474 | 5.476 | 23.000 | 2.260 | 118.01 (Sn-O-Sn) |
| SnO(β) | 2.744 | 90 | 90 | 90 | 5.474 | 5.476 | 23.000 | 2.256 | 118.09 (Sn-O-Sn) |

### 3.2 Band structure, density of states, and deformation charge density

Each O atom coordinates with four surface Sn atoms, forming a Sn–O tetrahedral structure. Due to the unique band structure of SnO, the valence band maximum (VBM) primarily consists of the 5*s* orbital of Sn and the 2*p* orbital of O. The lone electron pair formed by the 5*s* orbital of Sn points toward the interlayer spacing, resulting in a strong dipole-dipole interaction between adjacent SnO layers. The band structure of SnO, calculated using DFT along the special wave vector symmetry points in the Brillouin zone, is shown in **Fig. 3**. Fermi level $E_f$ is located at the position where the energy is 0. The band structure revealed that the conduction band lies above the Fermi surface, while the valence band resides below it. The highest point of the valence band and the lowest point of the conduction band are both located at the G point in the Brillouin zone. These points are in close proximity and distributed at both ends of the Fermi surface. Additionally, the valence band lies closer to the Fermi level, with one of its lines coinciding with the Fermi level. Therefore, it can be inferred that SnO is a direct narrow-bandgap semiconductor. The calculated results further reveal that the bandgaps for SnO(α) and SnO(β) are 1.153 eV and 1.090 eV, respectively. Meanwhile, the electron distribution pattern within the Brillouin zone, as shown in **Fig. 3**, reveals that the bandgap between the F and Q regions is the largest. The band structure in the F to Q region exhibits a relatively large bandgap with closely aligned energy level lines, indicating that SnO possesses excellent electron transport properties and good stability in this direction.

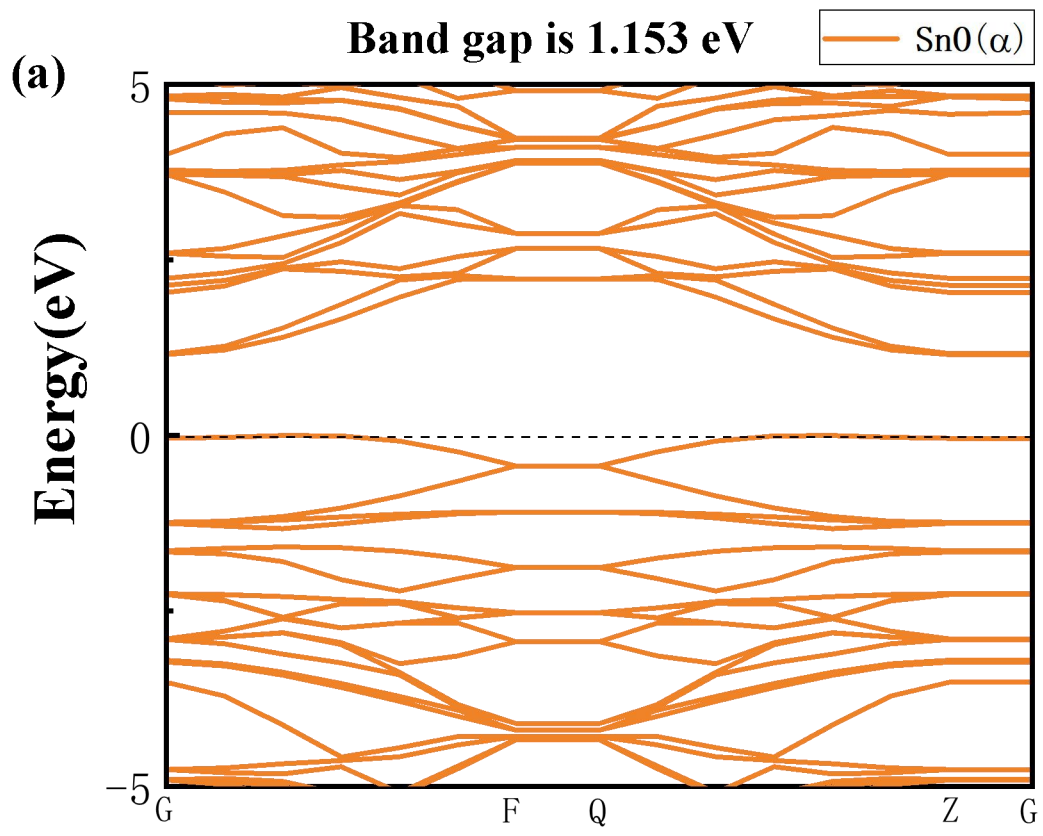
(a)
Band gap is 1.153 eV
Sn0(α)
Energy(eV)
5
0
−5
G
F
Q
Z
G

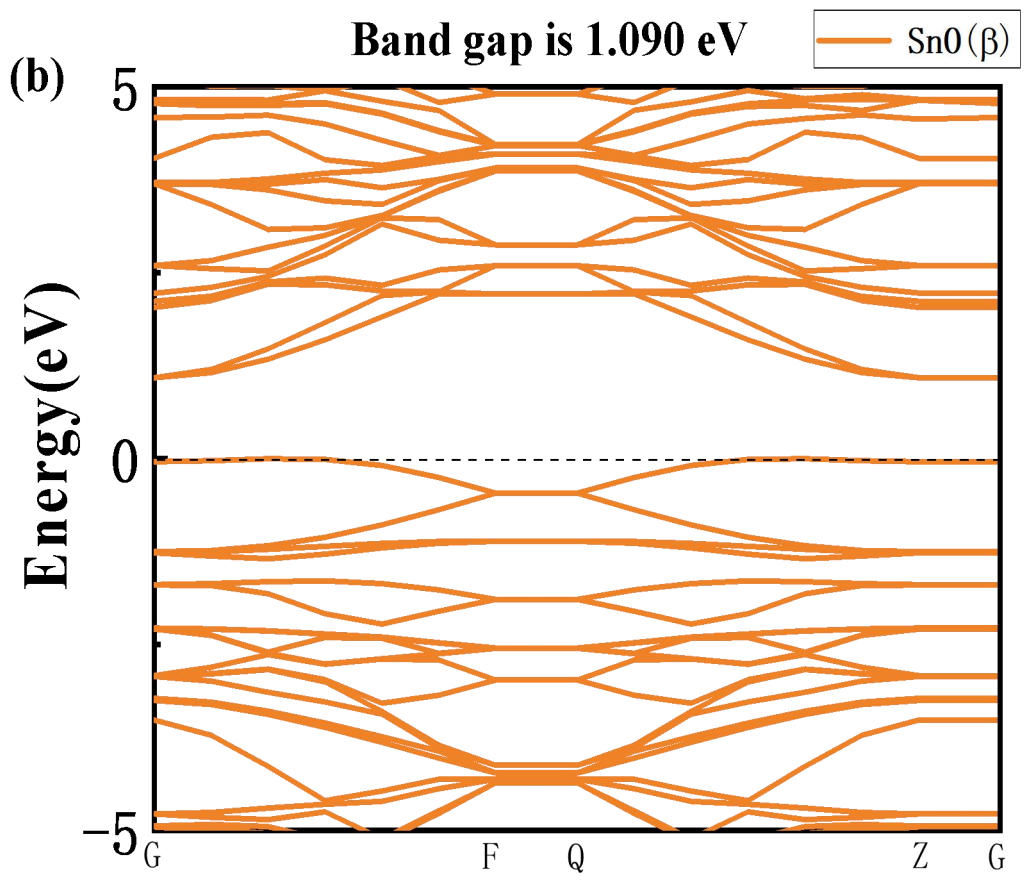
(b)
Band gap is 1.090 eV
Sn0(β)
Energy(eV)
5
0
−5
G
F
Q
Z
G

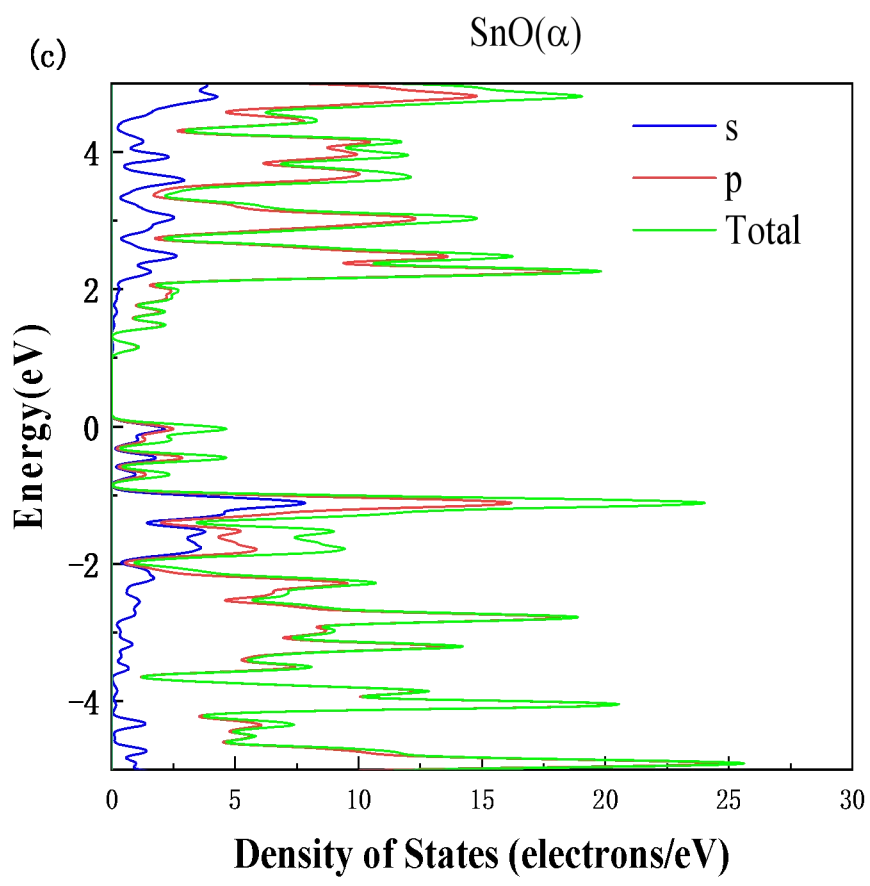
(c)
SnO(α)
s
p
Total
Energy(eV)
4
2
0
-2
-4
0
5
10
15
20
25
30
Density of States (electrons/eV)

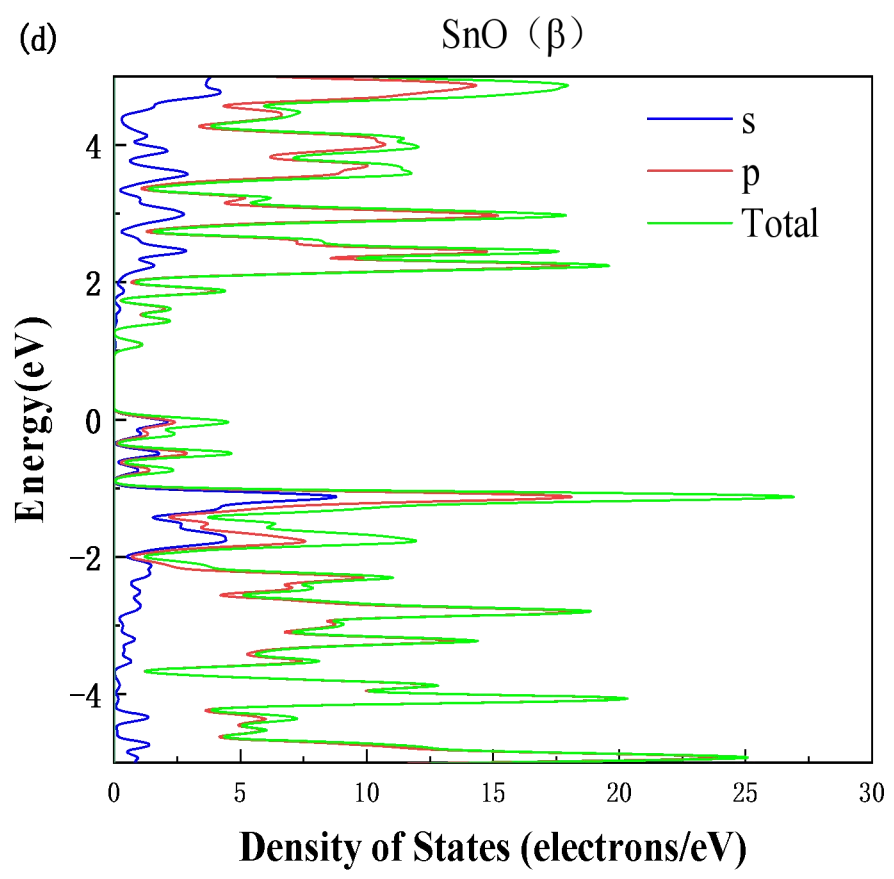

**Fig. 3** Band Structure and density of states of SnO(α) and SnO(β)

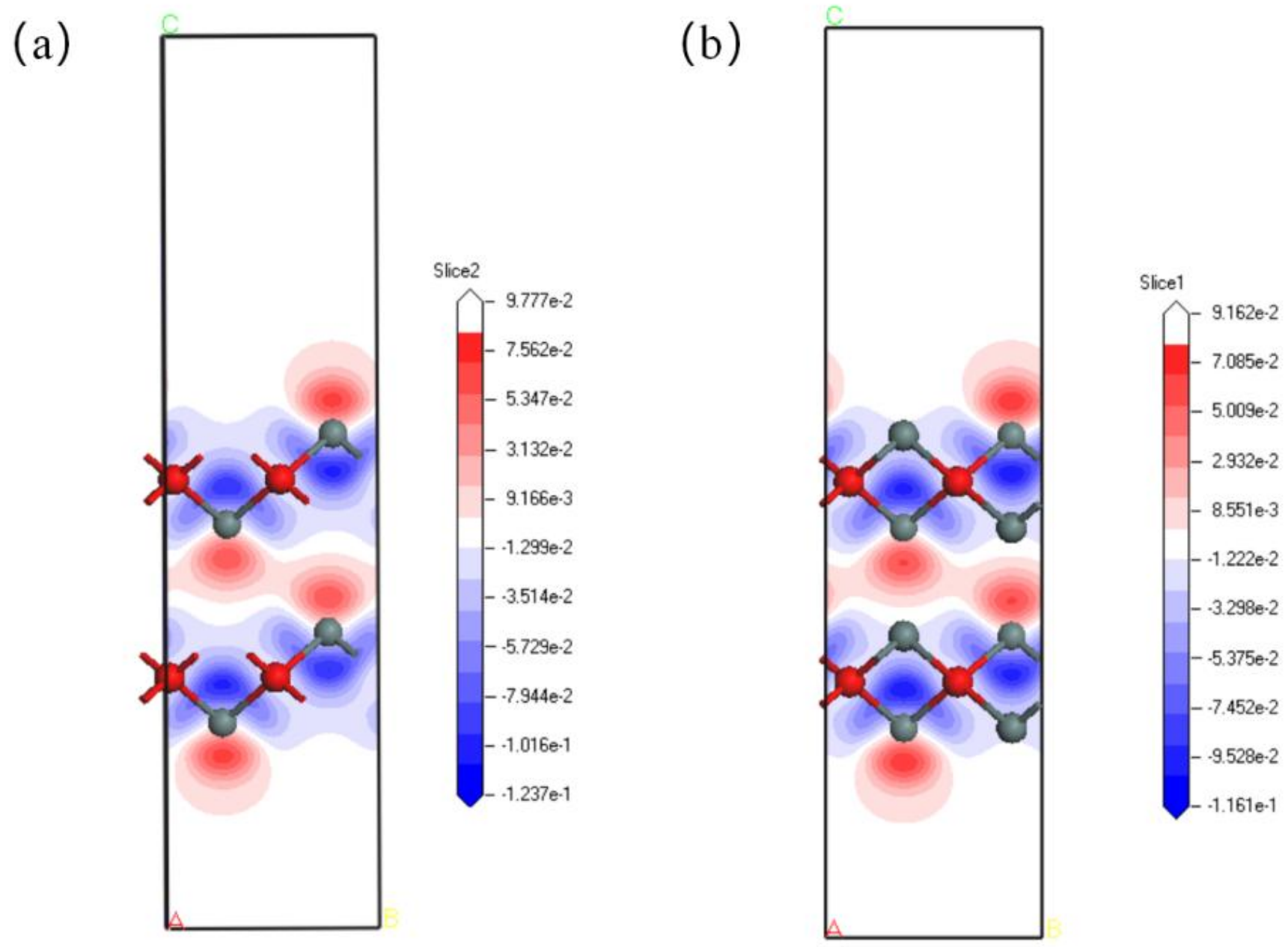

**Fig. 4** Deformation charge density of (a) SnO(α) and (b) SnO(β)

To explore the formation and distribution characteristics of the electronic energy band in SnO, the total and partial density of states (PDOS) of SnO were calculated. **Fig. 3** shows the calculated total and PDOS of SnO. In the total DOS, the Fermi level was within the range where the DOS value was zero, and the pseudogap was located near the Fermi level. In the PDOS diagram, it can be clearly seen that the PDOS of the *p*-orbital exhibits significant pseudogap characteristics, and the unique electronic structure corresponding to SnO is coupled with its p-type conduction properties.

Deformation charge density is an important parameter for studying electronic structures. This value was obtained by subtracting the atomic charge density at the corresponding points from the charge density after bonding. Through the calculation and analysis of the deformation charge density, we can clearly understand the migration of charges and properties such as the bonding polarization direction during the bonding and bonding electron coupling processes. In **Fig. 4**, blue represents regions with electron depletion, red represents regions with electron accumulation, and white represents regions where the electron density hardly changes. As shown in **Fig. 4**, Sn atoms surrounded by blue indicate electron loss during the bonding process, while O atoms surrounded by red indicate electron gain. Among them, the reddest regions of SnO(α) and SnO(β) gain $7.562\times10^{-1}$ electrons and $7.085\times10^{-1}$ electrons, respectively, and the bluest regions lose $-1.237\times10^{-1}$ electrons and $-1.161\times10^{-1}$ electrons, respectively (negative signs indicate electron loss).

This behavior is primarily due to the relatively low formation energy of Sn vacancies, as well as the more dispersed hole transport channels. The VBM hybrid energy level is formed by the hybridization of O 2*p* and spherical Sn 5*s* orbitals. The Sn 5*s* orbital plays a dominant role in the formation of the VBM. The Sn 5*s* orbital, being spherical with weak directionality of electron cloud overlap, facilitates hole transport. Therefore, SnO exhibits a relatively high hole mobility. According to **Fig. 3(a)**, the conduction band minimum energy level of SnO mainly originates from the Sn 5*p* orbital, and its PDOS curve exhibits a free electron band, indicating that if there are a sufficient number of electrons, electrons can also move in SnO to generate an electric current. These results provide an important theoretical basis and experimental support to further understand the electronic properties of SnO and its applications in related electronic devices.

The deformation charge density and electronic radius were used to calculate the deformation bond energies using **Eq. 15** as follows: -0.0675 and -0.0595 eV for SnO(α) and SnO(β), respectively. The parameters of the deformation charge density and atomic radius for the deformation bond energy for SnO(α) and SnO(β) obtained from the calculations are shown in **Table 3**.

**Table 3** Deformation charge density $\delta\rho\left(\overrightarrow{r_{ij}}\right)$ and deformation bond energy

$\delta V_{bc}\left(\vec{r_{ij}}\right)$, as obtained from the BBC model.

$\left(\varepsilon_0 = 8.85\times10^{-12} C^2 N^{-1} m^{-2}, e = 1.60\times10^{-19} C, \left|\vec{r_{ij}}\right| \approx d_{ij}/2\right)$

| | SnO(α) | SnO(β) |
|---|---|---|
| $r_{ij}$ ( Å ) | 1.130<br>Sn-O | 1.128<br>Sn-O |
| $r_i$ ( Å ) | 1.58<br>(Sn) | 1.58<br>(Sn) |
| $\vec{r}_j$ ( Å ) | 0.66<br>(O) | 0.66<br>(O) |
| $\delta\rho^{Hole-electron}\left(\vec{r_{ij}}\right)\left(e/\text{Å}^3\right)$ | -0.1237 | -0.1161 |
| $\delta\rho^{Bonding-electron}\left(\vec{r_{ij}}\right)\left(e/\text{Å}^3\right)$ | 0.0756 | 0.07085 |
| $\delta V_{bc}^{bonding}\left(\vec{r_{ij}}\right)(\text{eV})$ | -0.0675 | -0.0595 |

### 3.3 Analytical approach of the renormalization group on energy scales

An energy scale $L$ can be represented bond energy in the following way:

$$L = \left|\frac{V_i\left(\vec{r}\right)}{V_0\left(\vec{r}\right)}\right| = \mathrm{e}^{-\mu r} \tag{20}$$

Where $V_0\left(\vec{r}\right) = \left|\delta V_{bc}^{bonding}\left(\vec{r_{ij}}\right)\right|$ and $\mu$ is the charge-shielding factor [20]. The ultimate goal of studying energy scales is to resolve corresponding physical phenomena. The energy scale serves as a measure of bond energy, which determines the relaxation and strength of the energy. It can be expressed in the form of mathematical functions, whether linear or nonlinear, and can adapt to various physical phenomena.

The potential energy surface function of bonds can be represented by wave functions:

$$V_i\left(\vec{r}\right) = L\bullet V_0\left(\vec{r}\right)(\cos k_x + \cos k_y) = V_0\left(\vec{r}\right)\mathrm{e}^{-\mu r}(\cos k_x + \cos k_y) \tag{21}$$

**Eq. 21** can be used to plot the wave function diagram of the energy scale, as shown in **Fig. 5**. The wave function of the energy itself contains extremely rich physical information, displaying distinct characteristics at different scales. On a large scale, the image of the energy wave function can display macroscopic features such as the overall energy trend; for example, the overall energy distribution of a quantum system. At a small scale, it provides subtle fluctuations and local changes, such as slight differences in the particle probability distribution within a local area. The renormalization model effectively captures this multiscale characteristic and determines a balance between different scales through appropriate processing, thus enabling the effective transformation of the energy function.

To study the energy wave functions of SnO(α) $V_0\left(\vec{r}\right)=0.0675$, two different energy scale values were considered, $L$=2 and $L$=8, and these values were substituted into **Eq. 21** using the corresponding values from **Table 3**. The resulting formulas are as follows:

$$\begin{cases} V_i\left(\vec{r}\right)=0.0675\bullet(\cos k_x+\cos k_y) \\ V_i\left(\vec{r}\right)=2\bullet 0.0675\bullet(\cos k_x+\cos k_y) \\ V_i\left(\vec{r}\right)=8\bullet 0.0675\bullet(\cos k_x+\cos k_y) \end{cases} \tag{22}$$

For SnO(β)( $V_0\left(\vec{r}\right)=0.0595$ ), the study can also be conducted analogously with **Eq. 22**. Using these calculations, we obtained the wave function diagrams shown in **Fig. 5**, which illustrate the dynamic changes in the potential energy functions at different energy scales. In the 3D case (**Fig. 5a-f**), as $L$ increases, the system can cover a larger spatial range, and the variations in the wave function become more refined.

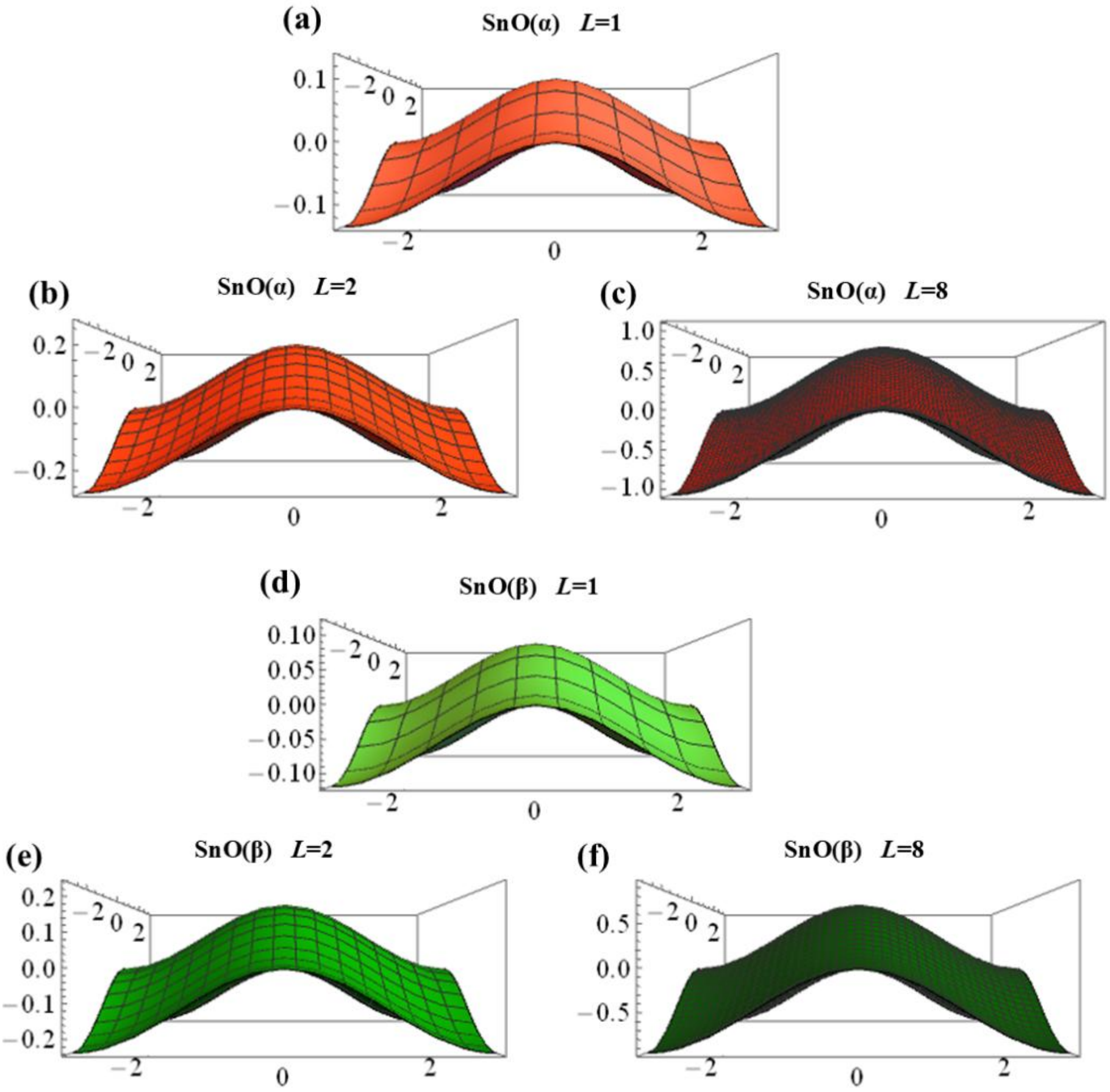

**Fig. 5** Renormalized energy scale wave function diagrams for SnO(α) and SnO(β):

(a) (b) (c) Respectively illustrate the renormalization of SnO (α) in 3D space when $L$ equals 1, 2, and 8; (d) (e) (f) The description pertains to the normalization of B in space under the same conditions.

Starting from **Eq. 10**, given that $b<1$, when $(n/2)\ (d-2)\ -d>0>0$, the coupling constant $\lambda_n$ decreases as the scale decreases, indicating that certain interactions become less significant at smaller scales and can thus be considered “irrelevant.” Conversely, when $(n/2)\ (d-2)\ -d>0<0$, the coupling constant $\lambda_n$ increases as the scale decreases, and these interactions are referred to as “relevant.” When $(n/2)\ (d-2)-d=0$, the coupling constant $\lambda_n$ does not change with scale, a situation defined as “marginal.” In scalar field theory, the most relevant interactions become marginal at the dimension known as the critical dimension in condensed matter physics. By taking the logarithms of both sides of **Eq. 10**, we can obtain

$$\ln(\lambda_n') = \left[\frac{n}{2}(d-2) - d\right]\ln(b) + \ln(\lambda_n)$$

（23）

Differentiating the above equation with respect to the scale $L$, and based on $b = 1 - (\delta\Lambda / \Lambda)$, with the scale $L = \Lambda^{-1}$, we obtain $\delta\Lambda = -\Lambda^2 dL$, hence

$$L\frac{d\lambda_n}{dL} = -\left[\frac{n}{2}(d-2) - d\right]\lambda_n$$

（24）

We chose the parameters of interacting fields n = 2, n = 6, and n = 8, and spatial dimensions d = 2 and d = 3 as subjects of our study and substituted their values into **Eq. 24** to explore the variation trend of the coupling constant $\lambda_n$ with the energy scale $L$.

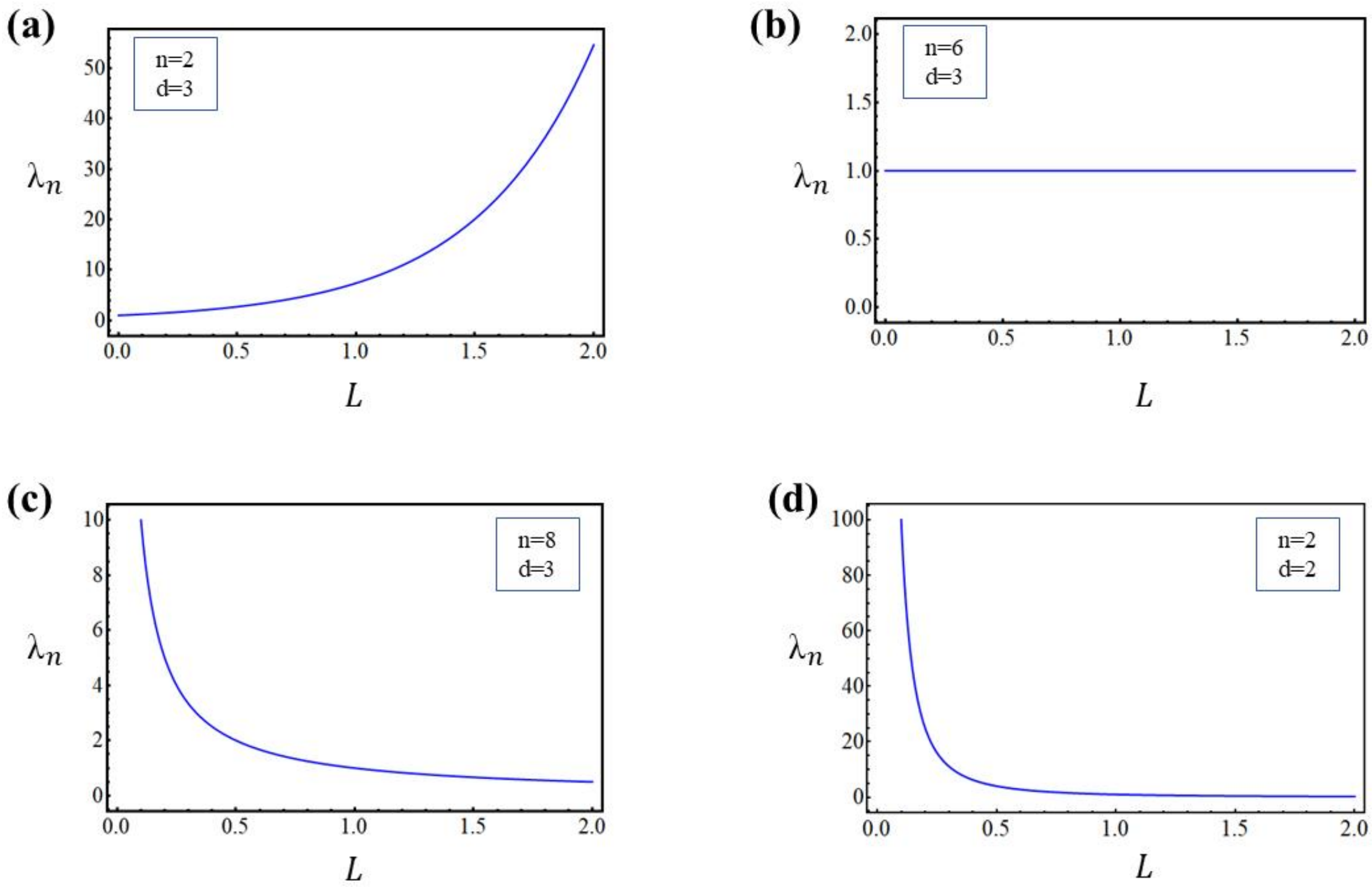

**Fig. 6** Coupling constant versus the energy scale. (a) Illustrates the case of n=2 in 3D space;(b) Describes the situation of n=6 in 3D space;(c) Represents the case of n=8 in 3D space;(d) Explains the scenario of n=2 in 2D space.

**Fig. 6** demonstrates the relationship between the energy scale $L$ and the coupling constant $\lambda_n$ in spaces of different dimensions. In 3D space when $n$=2 (**Fig. 6a**), the coupling constant $\lambda_n$ grows exponentially with the increase of the energy scale *L*, a phenomenon indicating that the interaction is of relatively low importance in renormalization group analysis. **Fig. 6b** then illustrates the marginal behavior at *n*=6, where the constancy of the coupling constant $\lambda_n$ is likely closely related to critical phenomena. In **Fig. 6c**, it can be observed that in 3D space with *n*=8, the coupling constant $\lambda_n$ decays exponentially as *L* increases, emphasizing the pivotal role of this interaction within the renormalization group framework. **Fig. 6d** describes the scenario in 2D space at *n*=2, where the coupling constant $\lambda_n$ decreases rapidly with the increase of *L*, similar to the case in 3D space at *n*=2, further highlighting the secondary role of this interaction in renormalization group analysis.

The renormalization group method is primarily used to deal with the behavior of physical systems at different scales. In physical systems, various interactions are described by coupling constants. When we consider the changes in the system at different scales—such as transitioning from the microscale to the macroscale—some system details may become unimportant, while certain overall and key physical properties must be retained.

The renormalization group method plays a crucial role in theoretical physics research, particularly in the discussion of critical phenomena and phase transitions. This method enables the analysis of the behavior of a system at different scales, particularly when the system is close to a critical point (quantum resolution limit). Under this framework, coupling constants are used to describe the strength of the interactions within the system and their behavior as the scale changes can be described using differential equations. The key point is that even if scale *L* is compressed, the main physical characteristics of the function can be preserved. This ensures that further analysis of scale *L*—such as calculating its energy eigenvalues and studying its correlation with other physical quantities—can still be performed accurately using the compressed function. During the compression process, we filtered and retained information that played a key role in the overall physical properties through renormalization operations.

## 4. Conclusion

In this study, the electronic and bonding properties of 2D SnO materials were calculated using DFT. The calculations and analysis based on the BBC model provided clear insights into the topological geometry of atomic bonds and the state of bonding electrons. In addition, the energy wavefunction was renormalized to determine the energy wavefunction at varying quantum resolution sizes. The results revealed that the coupling constant $\lambda_n$ changes with scale $L$. This study greatly improves our understanding of the local bonding state and quantum resolution size on the surfaces of 2D structural materials.

## Acknowledgment

Financial support was provided by the Science and Technology Research Program of the Chongqing Municipal Education Commission (Grant No. KJQN202401420).

# Supplemental Material

## Quantum resolution sizes and atomic bonding states of two-dimensional SnO

**Yu Wang, Yunhu Zhu,Yixin Li, Maolin Bo***

Key Laboratory of Extraordinary Bond Engineering and Advanced Materials Technology (EBEAM) of Chongqing, Yangtze Normal University, Chongqing 408100, China

*Author to whom correspondence should be addressed: bmlwd@yznu.edu.cn (Maolin Bo)

## 1.Hamiltonian

For bond-charge (BC) model, we consider the positive charge background (b) and the electron (e) as a system, and write their Hamiltonian sums and their interactions, respectively. In addition to electron kinetic energy, only electrostatic Coulomb interactions are considered:

$$H = H_b + H_e + H_{eb} \tag{1}$$

$$H_e = \sum_{i=1}^{N} \frac{P_i^2}{2m} + \frac{1}{2} e_1^2 \sum_{i=1}^{N} \sum_{\substack{j=1 \\ i \neq j}}^{N} \frac{1}{|r_i - r_j|} e^{-\mu|r_i - r_j|} \tag{2}$$

$$H_b = \frac{1}{2} e_1^2 \int \mathrm{d}^3 x \int \mathrm{d}^3 x' \frac{n(x) n(x')}{|x - x'|} e^{-\mu|x - x'|} \tag{3}$$

$$H_{eb} = -e_1^2 \sum_{i=1}^{N} \int \mathrm{d}^3 x \frac{n(x)}{|x - r_i|} e^{-\mu|x - r_i|} = -e_1^2 \sum_{i=1}^{N} \frac{N}{V} 4\pi \int \mathrm{d}z \frac{e^{-\mu z}}{z} = -4\pi e_1^2 \frac{N^2}{V\mu^2} \tag{4}$$

The shielding factor $\mu$ is added to the equation. $r_i$ represents the *ith* electronic position. $x$ represents the background position. $e$ is the basic charge, $e_1 = e/\sqrt{4\pi\varepsilon_0}$. Since the coulomb force is a long-range force, while $N \to \infty$, $V \to \infty$, $N/V$ remains unchanged, if the coulomb potential energy is calculated separately, $H_e$、$H_b$ and $H_{eb}$ will diverge. In physically, when increasing the volume by keeping the electron density ($N/V$) constant, the average electron energy $H/N$ should be a constant. Therefore, the divergent terms in the $H_e$, $H_b$ and $H_{eb}$ should be able to cancel each other out, in order to see this, We have added an attenuation factor $\mu$ to (2), (3), and (4), after the integration $L$ of the cube with a side length of $L \to \infty$, And then order $\mu \rightarrow 0$ , this is a commonly used technique in solid state physics calculations. $e_1 n(x)$ is the charge density at background $x$ , and $n(x) = N/V$ is a constant.

$$H_b = \frac{1}{2} e_1^2 \left(\frac{N}{V}\right)^2 \int \mathrm{d}^3 x 4\pi \int \mathrm{d}z \frac{e^{-\mu z}}{z} = 4\pi e_1^2 \frac{N^2}{2V\mu^2}$$

（5）

$$H_{eb} = -e_1^2 \sum_{i=1}^{N} \frac{N}{V} 4\pi \int \mathrm{d}z \frac{e^{-\mu z}}{z} = -4\pi e_1^2 \frac{N^2}{V\mu^2}$$

（6）

$$\sum_{i=1}^{N} \frac{P^2}{2m} = \sum_{l'\sigma'} \sum_{l\sigma} a_{k_{l'}\sigma'}^{\dagger} \left\langle k_{l'}\sigma' \left| \frac{P^2}{2m} \right| k_l \sigma \right\rangle = \sum_{l\sigma} \frac{\hbar^2 k_l^2}{2m} a_{k_l\sigma}^{\dagger} a_{k_l\sigma}$$

(7)

In the above formula, use $k_l$ representative $k_{xl}$ 、 $k_{yl}$ and $k_{zl}$ , $l = 0, \pm 1, \pm 2, \pm 3, \cdots$ for typographical convenience, we sometimes use $k$ and $k'$ in lieu of $k_l$ and $k_{l'}$ , just remember the momentum $k$ take discrete values instead of continuous values. The second item in（2） is

$$\begin{aligned} &\frac{1}{2} e_1^2 \sum_{l'\sigma'} \sum_{m'\lambda'} \sum_{l\sigma} \sum_{m\lambda} a_{k_{l'}\sigma'}^{?} a_{k_{m'}\lambda'}^{\dagger} (k_{l'}\sigma', k_{m'}\lambda' \left| \frac{e^{-\mu|r_1 - r_2|}}{|r_1 - r_2|} \right| k_l\sigma, k_m\lambda) a_{k_m\lambda} a_{k_l\sigma} \\ &= \frac{1}{2} e_1^2 \sum_{l'} \sum_{m'} \sum_{l\sigma} \sum_{m\lambda} \delta_{\sigma\sigma'} \delta_{\lambda\lambda'} a_{k_{l'}}^{?} a_{k_{m'}}^{\dagger} (k_{l'} k_{m'} \left| \frac{e^{-\mu|r_1 - r_2|}}{|r_1 - r_2|} \right| k_l k_m) a_{k_m} a_{k_l} \end{aligned}$$

（8）

The matrix elements are first calculated in the position representation. Insert two completeness relations in the matrix element in the above equation:

$$\int d^3 r^{\alpha} \int d^3 r^{\beta} \left| r^{\alpha} r^{\beta} \right\rangle \left\langle r^{\alpha} r^{\beta} \right| = 1$$
$$\int d^3 r^{\gamma} \int d^3 r^{\delta} \left| r^{\gamma} r^{\delta} \right\rangle \left\langle r^{\gamma} r^{\delta} \right| = 1$$

And

$$(k_{l'}k_{m'}\left|\frac{e^{-\mu|r_1-r_2|}}{|r_1-r_2|}\right|k_lk_m) = \frac{1}{V^3}\int d^3r^\alpha \int d^3r^\beta e^{-i(k_{l'}-k_l)r^\alpha} e^{-i(k_{m'}-k_m)r^\beta} \frac{e^{-\mu|r^\alpha-r^\beta|}}{|r^\alpha-r^\beta|}$$

$$= \delta\left(k_{l'}+k_{m'},k_l+k_m\right)\frac{1}{V}\int \frac{e^{-\mu r}}{r}e^{-i(k_{l'}-k_l)r}d^3r$$

$$= \delta\left(k_{l'}+k_{m'},k_l+k_m\right)\frac{2\pi}{V}\int re^{-\mu r}\left(\frac{2}{kr}\sin kr\right)dr$$

$$= \delta\left(k_{l'}+k_{m'},k_l+k_m\right)\frac{1}{V}\frac{4\pi}{\left(k_{l'}-k_l\right)^2+\mu^2}$$

（9）

In the above calculation, the second step uses the independent variable substitution： $r^\alpha + r^\beta = r'$ 、 $r^\alpha - r^\beta = r$ [Jacobian determinant $J=(1/2)^3$ ]，substitute this equation back（8），$k_l = k$ ，$k_m = k'$ ，$k_{l'} = k+q$ ，$k_{m'} = k'-q$ ，the second item in $H_e$ is

$$\frac{e_1^2}{2V}\sum_k\sum_{k'}\sum_q\sum_{\sigma\lambda}\frac{4\pi}{q^2+\mu^2}a^\dagger_{k+q,\sigma}a^\dagger_{k'-q,\lambda}a_{k'\lambda}a_{k\sigma}$$

（10）

Separate out $q=0$ is

$$\frac{e_1^2}{2V}\sum_k\sum_{k'}\sum_q\sum_{\sigma\lambda}\frac{4\pi}{\mu^2}a^\dagger_{k\sigma}a^\dagger_{k'\lambda}a_{k'\lambda}a_{k\sigma} = \frac{e_1^2}{2V}\frac{4\pi}{\mu^2}\left(N^2-N\right)$$

*N* is the total number of particles，because

$$a^\dagger_{k\sigma}a^\dagger_{k'\lambda}a_{k'\lambda}a_{k\sigma} = N_{k\sigma}N_{k'\lambda} - N_{k\sigma}\delta_{kk'}\delta_{\sigma\lambda}$$

Now substitute (7) and (10) back to equation (2), $H_e$ is

$$H_e = \sum_{k\sigma}\frac{\hbar^2k^2}{2m}a^\dagger_{k\sigma}a_{k\sigma} + \frac{e_1^2}{2V}\sum_q{}^*\sum_{k\sigma}\sum_{k'\lambda}\frac{4\pi}{q^2+\mu^2}a^\dagger_{k+q,\sigma}a^\dagger_{k'-q,\lambda}a_{k'\lambda}a_{k\sigma} + \frac{e_1^2}{2V}\frac{4\pi}{\mu^2}\left(N^2-N\right)$$

（11）

In the formula, the *q* on the sum sign of "*" indicates that the part where q=0 is ignored during the sum. When $V\to\infty, N\to\infty$, while keeping *N*/*V* constant, the last term to the right of the equal sign causes the average energy $H_e/N$ of each particle to become:

$$\frac{1}{2}4\pi e_1^2\left(\frac{N}{V}\right)\frac{1}{\mu^2}-\frac{1}{2}4\pi e_1^2\left(\frac{N}{V}\right)\frac{1}{N}\frac{1}{\mu^2}$$

The former term is constant, and the latter term tends to zero. If $\mu \to 0$, the former term becomes a divergent term. However, this term just cancels out the divergent $H_b$ and $H_{eb}$ term. Thus, the Hamiltonian of the system becomes

$$H=\sum_{k\sigma}\frac{\hbar^2k^2}{2m}a_{k\sigma}^{\dagger}a_{k\sigma}+\frac{e_1^2}{2V}\sum_{q}\sum_{k'}\sum_{\sigma\lambda}\frac{4\pi}{q^2}a_{k+q,\sigma}^{\dagger}a_{k'-q,\lambda}^{\dagger}a_{k'\lambda}a_{k\sigma}=H_0+H_1 \tag{12}$$

**Structural stability**

We used molecular dynamics (MD) to simulate the structural stability of SnO. The SnO(α) and SnO(β) structure were obtained by performing NVE ensemble with time increments at 1 fs for 100 ps (the total iteration steps are 100,000) until the potential energy accomplished a stable value. The box size and volume of SnO(α) and SnO(β) , as is shown in the **Table S1**. The molecular dynamics simulation results show that the structures of SnO(α) and SnO(β) are stable.

**Table S1** Box size, volume and number of atoms of SnO

| Structure | Volume | Box size | | | Angle | | |
|---|---|---|---|---|---|---|---|
| | | a(Å) | b(Å) | c(Å) | α(°) | β(°) | γ(°) |
| SnO(α) | 24817.169 Å$^3$ | 32.8429 | 32.44 | 23.000 | 90.000 | 90.000 | 90.000 |
| SnO(β) | 24817.169Å$^3$ | 32.8429 | 32.8429 | 23.000 | 90.000 | 90.000 | 90.000 |

**2.Molecular dynamics (MD) to simulate the structural stability of SnO(α) and SnO(β)**

(1) Structural stability of SnO(α)

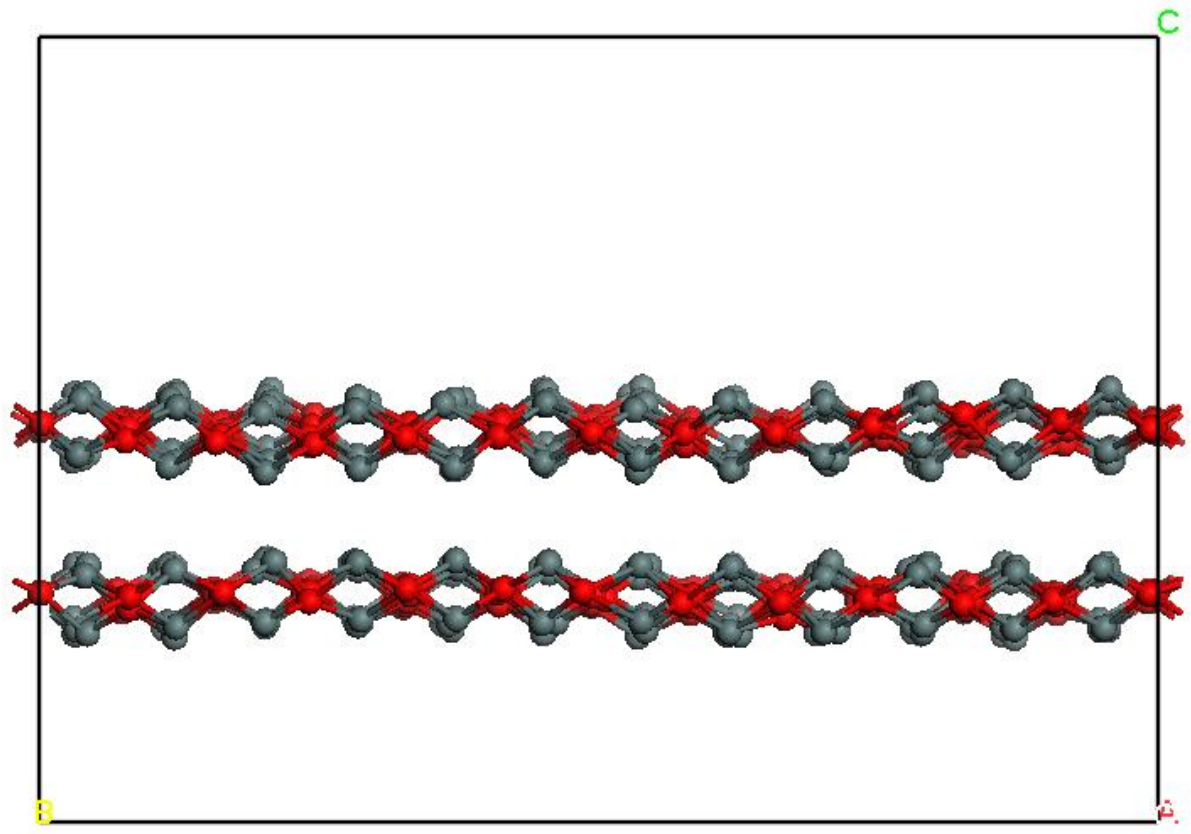

298K

(2) Structural stability of SnO(β)

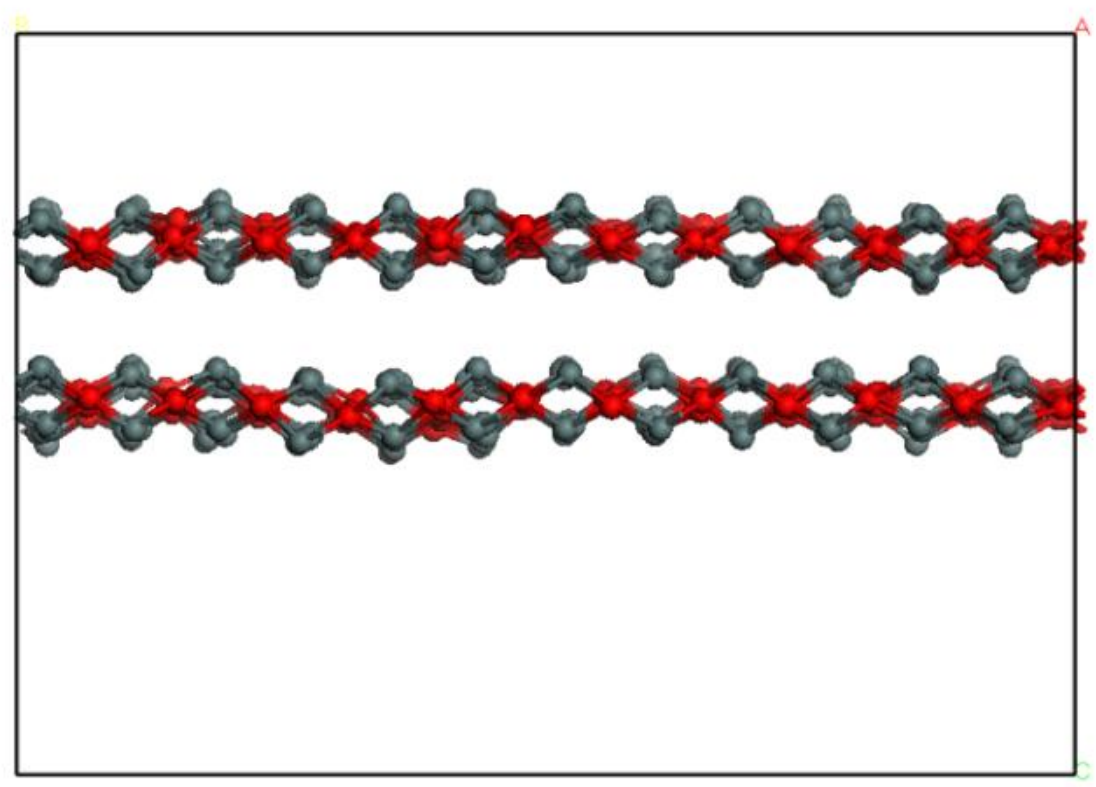

298K